\documentstyle[aps,prl,multicol,amssymb,epsf]{revtex}
\title{Purifying Noisy Entanglement Requires Collective 
Measurements}

\author{N. Linden$^1$, S. Massar$^2$ and S. Popescu$^{1,3}$}

\address{
$^1$Isaac Newton Institute for Mathematical
Sciences, Cambridge, CB3 0EH, UK\\%
${}^2$Institute for Theoretical Physics,
      Princetonplein 5, PO Box 80006, 3508 TA Utrecht, The 
Netherlands \\
$^3$BRIMS, Hewlett-Packard Laboratories, Stoke
Gifford, Bristol BS12 6QZ, UK}

\date{30 April 98}
\begin{document}
\draft
\maketitle
\begin{abstract}
Known entanglement purification protocols for mixed states use
collective measurements on several copies of the state in order to
increase the entanglement of some of them. We address the question of whether 
it 
is possible to purify
the entanglement of a state by processing each copy separately.  While 
this is possible for pure states, we show that this is impossible, in general,
for mixed states.
The importance of this result both conceptually and for experimental 
realization 
of
purification is discussed. We also give explicit invariants of an
entangled  state of two qubits under local actions and classical
communication. 
\end{abstract}


\begin{multicols}{2}

%%% Mathematical odds and ends

\newcommand\mathC{\mkern1mu\raise2.2pt\hbox{$\scriptscriptstyle|$}
                {\mkern-7mu\rm C}}		       
\newcommand{\mathR}{{\rm I\! R}}                % The real numbers 

%%%

Entanglement is perhaps  the key  resource which distinguishes
quantum from classical information theory. It
plays a central role in quantum computation \cite{B} and
quantum error correcting codes\cite{BDSW}, and it gives rise to some
completely new applications such as 
dense coding \cite{BW}, teleportation \cite{BBCJPW} and
certain forms of quantum cryptography \cite{E}.
In order to function optimally these applications require 
maximally entangled states. Otherwise the dense coding, teleportation
or quantum cryptography will be imperfect/noisy.
However interactions 
with the environment always occur, and will degrade the quality of
the entanglement.  
If the destructive effects of the environment are
not too important, then they can be counteracted by 
``entanglement purification''\cite{BBPS,BBPSSW,G2,H3}. 
This is realized by
 carrying out local measurements on the entangled 
particles and classical communication. The entanglement of some pairs 
is thereby increased at the expense
of the others which are destroyed.

There are two fundamentally different types of purification protocols: those 
acting on individual pairs of entangled particles and those acting collectively 
on many pairs.  In this letter we address the following question: 

{\em Is it the case that, 
whenever
it is possible to purify by collective actions, it is also 
possible to purify by actions on individual pairs?}  

In the case of pure states this is indeed true.
One can always, with finite probability, bring an individual entangled
 pure state to a maximally entangled state using only local 
operations\cite{BBPS}.
The main result of this letter is to show that there are situations in which 
entanglement cannot be  purified by actions on 
individual pairs, even though it 
can be purified by collective actions.
  This result is  
surprising because
we  expect entanglement to be a property of each pair
 individually rather than a global property of
many pairs.

Specifically we consider the case of Werner density matrices \cite{W} for 
two spin-1/2 particles.  It is known \cite{BBPSSW} that it is always 
possible to purify singlets from Werner density matrices by collective methods
(if the initial Werner density matrix is entangled).
However we will show that is not possible to purify singlets, or even 
increase the fidelity of a Werner density matrix infinitesimally,
by any 
combination of local actions and classical communication acting on  individual 
pairs. 
This is the case even though Werner states {\em do} have active 
non-locality at the single-pair level, since a single Werner state can realize 
teleportation (although the teleportation is
imperfect, it nonetheless has fidelity
better than any classical procedure\cite{P2}).

As well as its implications for conceptual aspects of non-locality, our result 
has relevance to the experimental realization of purification.
The
main experimental difficulty, which has so far prevented implementation in the 
laboratory,
is  that purification
protocols generally 
require collective measurements on many entangled pairs. Such
measurements are very delicate as they involve 
controlled  interactions among different particles. 
On the other hand measurements on individual particles are much easier
to realize. For instance photo-detectors and 
polarized beam splitters 
efficiently realize von Neumann
measurements on the polarization of photons. 
More general Positive Operator
Valued Measurements (POVM's) which necessitate the use of an ancilla have 
already 
been carried out. 
In the case  of photon polarization, the momentum of the photon
serves as a convenient ancilla and an arbitrary POVM on a photon can be
realized with present technology, see \cite{G} and
\cite{P,BBDHP}.

Thus our result is disappointing from an experimental point of view
since it means that purification of arbitrary states cannot be realized
using present technology.

We now turn to the proof of our result.
We consider Werner
states \cite{W}, namely states of the form
\begin{equation}
W(F) = F S + {1 - F \over 3}(1_4 - S)
\end{equation}
where $S$ is the projection operator onto the singlet state,
 $\psi = (\uparrow\downarrow - \downarrow\uparrow ) / \sqrt{2}$ and
$1_4$ is the $4\times 4$ identity matrix. 
$F =
 {\rm tr} (   W(F) S )$ is the fidelity of the
 Werner state. These
 states play a central role in purification protocols, because by
 carrying out suitably chosen unitary transformations on both particles, one 
can
 always bring any entangled state to the Werner form.
For $F\leq 1/2$ 
a Werner state is unentangled and can be expressed as a mixture of
product states.
But for $1>F> 1/2$ there are purification protocols which can extract
 states with arbitrary large entanglement from an initial set of Werner
 states. 
The simplest purification protocol which has been described uses
 collective measurements on pairs of Werner states\cite{BBPSSW}. 
We shall
 show that it is impossible to increase the fidelity of  a 
 Werner state by local operations and classical communication on an individual 
copy.

Consider a {\em  single} copy of the
mixture $\rho$ of two qubits shared between Alice and Bob
(later we will consider the specific case of a Werner state). After
carrying out local actions and classical communication they will
obtain a density matrix $\rho_{final}$. 
In our proof it will be convenient to use 
the ``entanglement of
formation''\cite{BDSW} as measure of  the
entanglement of $\rho$ and $\rho_{final}$.
It is defined as follows:
\begin{itemize}
\item For a pure state $|\psi>$ shared between Alice and Bob,
$E(\psi) = - {\rm tr}\rho_A \ln_2  \rho_A = - {\rm tr} \rho_B \ln_2  \rho_B$
where
$\rho_A = {\rm tr}_B |\psi><\psi|$ and $\rho_B = {\rm tr}_A |\psi><\psi|$.
\item For a mixed state $\rho$ the entanglement of formation is the
 minimum entanglement of the mixtures of pure states that realize $\rho$: 
$E(\rho) = \min \sum_i p_i E(\psi_i)$ where the minimum is taken over
 all $p_i$, $\psi_i$ such that 
$\rho = \sum_i p_i |\psi_i><\psi_i|$.
\end{itemize}

Hill and Wootters have given an explicit formula 
for the entanglement of formation in the case  
of two entangled qubits\cite{HW,WW}.
They introduce the operation of time reversal
$\tilde{\ }$. For single qubit the density matrix may be written as
$\rho = {1\over 2}(1_2 + {\bf
\alpha}. \sigma)$ (where ${\bf \alpha}. {\bf \alpha} \leq 1$, $1_2$ is the 
$2\times 2$ identity matrix,
and $\sigma_i$ are the Pauli matrices).
Then $\tilde \rho := \sigma_2 \rho^* \sigma_2 =
 {1\over 2}(1_2-  {\bf
\alpha}. \sigma)$, where complex conjugation is performed in the basis in which
$\sigma_z$ is diagonal. For a state of two qubits, the time reversal
operation is:
$\tilde{\rho } = \sigma_2 \otimes \sigma_2 \rho^*  \sigma_2 \otimes
\sigma_2$. 
Now consider the (non-Hermitian, but positive) matrix $\rho \tilde \rho$ and 
denote by
$\lambda_i$ the positive square root of its eigenvalues:
\begin{equation}
\rho \tilde \rho |\tilde v_i> = \lambda_i^2 |\tilde v_i>.
\end{equation}
The ``concurrence'' of the state $\rho$ is defined by
\begin{equation}
C(\rho) = \max \{0, \lambda_1 - \lambda_2 - \lambda_3 - \lambda_4\}
\end{equation}
where the $\lambda_i$ are taken in decreasing order, and the entanglement of 
formation $E(\rho)$ is 
\begin{eqnarray}
E(\rho) &=& H( {1 + \sqrt{ 1 - C^2(\rho)}\over 2})\nonumber\\
 \hbox{\rm where}\quad H(p) &=& - p \ln_2 p - (1-p) \ln_2 (1-p) 
\end{eqnarray}
Note that $ {E}(C)$ is a strictly monotonic function of $C$ so
that the concurrence is a measure of entanglement which is equivalent
to the entanglement of formation, ie. $E(\rho_1) = E(\rho_2)$
if and only if $C(\rho_1)= C(\rho_2)$.

To proceed we must recall what are the
possible local operations that can be carried out on a density matrix
$\rho$ and describe them explicitly. Then we shall compute
how the entanglement of formation changes under these local operations.
Consider a mixture $\rho$ of two qubits shared between Alice and Bob. Any
purification protocol can be conceived as successive rounds of
measurements and communication by Alice and Bob. Suppose Alice carries
out the first measurement. It can have many different
outcomes. Let us suppose that it has outcome $i_1$. Then after the
measurement  the state of
the system becomes $A_{i_1}\rho  A^\dagger_{i_1}$, up to normalization, 
where $A_{i_1}$ is an arbitrary operator (in general 
non-Hermitian) acting on the Hilbert space of Alice's
particle 
($A^\dagger_{i_1} A_{i_1}$ are the elements of the POVM realized
by Alice\cite{MEAS}).
After communicating the result of her measurement to Bob,
he carries out a measurement and obtains outcome $j_1$. The state of the
system is then $[A_{i_1}\otimes B_{j_1}(i_1)] \rho   [A^\dagger_{i_1}
\otimes B^\dagger_{j_1}(i_1)]$ where $B_{j_1}(i_1)$ is an arbitrary operator
acting on the Hilbert space of Bob's particle which can depend on the
outcome $i_1$ of Alice's measurement. Therefore after $N$
rounds of measurements and communication, the state of the system can
always be written as
\begin{equation}
\rho_{final} = 
{
A\otimes B \rho A^\dagger\otimes B^\dagger
\over
{\rm tr} (A\otimes B \rho A^\dagger\otimes B^\dagger)
}
\end{equation}
 where $A$ and $B$ are 
arbitrary operators acting on Alice's and Bob's Hilbert space
respectively. ($A$ denotes the product of the $N$ operators $A_{i_1}$, ...,
$A_{i_N}(i_1,j_1,i_2,...j_{N-1})$ representing the effects of the $N$
measurements carried out by Alice, and similarly for $B$).

We will need below an explicit expression for $A$ and $B$. To this end
note that 
we can always write an arbitrary operator $A$ in the form $ A = U_{A2} f_A 
U_{A1}$ 
where  $U_{A1}$ and $U_{A2}$ are unitary 
operators and
$f_A = \nu(1_2 + a \sigma_z)$, with $0\leq a \leq 1$ and $0<\nu\leq 1/(1+a)$,
is a filtration along the $z$ axis.  The upper bound on $\nu$ arises from the 
fact that, for $f_A$ to be physically realisable, its eigenvalues must be 
between zero and one.  
The filtration changes the relative weights of the
components of the spin along the $+z$ and $-z$  directions. 
We now write 
$A = U_{A2}U_{A1}U_{A1}^\dagger f_A^{a,{\bf z}} U_{A1} 
= U_A f_A^{a, {\bf n}}$ where $U_A =  U_{A2}U_{A1}$ and 
$f_A^{a, {\bf n}} =  \nu(1_2 + a {\bf n}.\sigma)$ and ${\bf n}$ is the
vector $+z$ rotated by the action of $U_{A1}$. This is the expression
we shall use below.

In addition to carrying out local measurements and communication,
Alice and Bob could also randomize the state they obtain. That is they
``forget'' which operations they carried out and thus obtain
a convex combinations of different final states
$\sum_i p_i \rho_{final}^i$. However such randomization can only
decrease 
the entanglement: $E(\sum_i p_i \rho_{final}^i) \leq \sum_i p_i
E(\rho_{final}^i)$, as shown in \cite{BBPSSW,BDSW}. This is natural since 
randomization 
loses
information about the state $\rho_{final}$.  For this reason we shall suppose 
that
Alice and Bob keep all the information available to them  and do not 
carry out randomization.

Having described how $\rho$ changes under local operations, we must
describe how $\tilde \rho$ changes. We will
then be in a position to calculate how the concurrence changes under
local  operations. Let us first collect some properties of the time
reversal operation.
\begin{itemize}
\item if $\rho = \rho_A \otimes \rho_B$, then
$\tilde \rho = \tilde \rho_A \otimes \tilde\rho_B$.
\item if $\rho = O \rho' O^\dagger$ where $O$ is a (possibly non-Hermitian) 
operator,  then
$ \tilde \rho = \tilde O \tilde \rho' \tilde O^\dagger$
\item
if $U_A = \cos\theta 1_2 + i \sin \theta {\bf q}. \sigma$ 
is a unitary transformation carried out by Alice, then
$\tilde U_A = U_A$.
\item
if $f_A^{a,{\bf n}}$ is a filtration carried out by Alice, then
$\tilde f_A^{a,{\bf n}} = f_A^{a,{\bf -n}}$.
\end{itemize}

Therefore since 
\begin{eqnarray}
\rho_{final} =
{U_A f_A^{a,{\bf n}}\otimes U_B f_B^{b,{\bf m}} \rho 
 f_A^{a,{\bf n}}  U_A^\dagger \otimes f_B^{b,{\bf m}}  U_B^\dagger
\over t(\rho; a , {\bf n} ; b,{\bf m}) },
\label{arbitrary}
\end{eqnarray}
with the normalization
\begin{eqnarray}
 & & t(\rho; a , {\bf n} ; b,{\bf m}) = 
{\rm tr} \left[
f_A^{a,{\bf n}} f_A^{a,{\bf n}}  \otimes f_B^{b,{\bf m}}f_B^{b,{\bf m}}
\rho \right],
\end{eqnarray}
then
\begin{eqnarray}
& &\tilde \rho_{final}=\nonumber\\ 
& &\quad {U_A f_A^{a,{\bf -n}}\otimes U_B f_B^{b,{\bf -m}} \tilde \rho 
 f_A^{a,{\bf -n}}  U_A^\dagger \otimes f_B^{b,{\bf -m}}  U_B^\dagger
\over t(\rho; a , {\bf n} ; b,{\bf m}) }.
\end{eqnarray}
We have defined $f_B^{b,{\bf m}}=\mu(1_2 + b {\bf m.\sigma})$, where
$0\leq b\leq 1$ and $0<\mu\leq 1/(1+b)$.

Using the fact that $f_A^{a,{\bf n}} f_A^{a,{\bf -n}} = \nu^2(1- a^2) 1_2$ and
$f_B^{b,{\bf m}} f_B^{b,{\bf -m}} = \mu^2(1- b^2) 1_2$,
one finds that 
\begin{eqnarray}
&\rho_{final}& \tilde \rho_{final}
=
{ \mu^2\nu^2(1-a^2)(1-b^2) \over t^2(\rho; a , {\bf n} ; b,{\bf m})}
\nonumber\\
&{}&\times U_A f_A^{a,{\bf n}}\otimes U_B f_B^{b,{\bf m}} \rho \tilde \rho 
 f_A^{a,{\bf -n}}  U_A^\dagger \otimes f_B^{b,{\bf -m}}  U_B^\dagger.
\end{eqnarray}
>From this expression one obtains the eigenvalues of 
$\rho_{final} \tilde \rho_{final}$ which we need to compute the
concurrence of $\rho_{final}$:
\begin{eqnarray}
&\rho_{final}& \tilde \rho_{final} |\tilde w_i> \nonumber\\
&=& 
{\mu^4\nu^4 (1-a^2)^2(1-b^2)^2 \over t^2(\rho; a , {\bf n} ; b,{\bf m})}
\lambda_i^2 |\tilde w_i> \end{eqnarray}
where
\begin{eqnarray}
|\tilde w_i> &=& U_A f_A^{a,{\bf n}}\otimes U_B f_B^{b,{\bf m}}|\tilde
v_i>
\end{eqnarray}
and $|\tilde v_i>$ is an eigenvector of $\rho \tilde \rho$
with eigenvalue $\lambda_i^2$. 
Hence 
\begin{equation}
C(\rho_{final})
= { \mu^2\nu^2(1-a^2)(1-b^2) \over t(\rho; a , {\bf n} ; b,{\bf m})}
C(\rho).
\end{equation}
Since the entanglement of formation is a strictly increasing
function of the concurrence $C(\rho)$, the entanglement of formation
can only increase if $C$ increases.

To complete the calculation we need the normalization $t$. To
this end we introduce the following representation of a density matrix
of two qubits 
\begin{equation}
\rho = {1 \over 4}[1_4 + {\bf \alpha}. \sigma\otimes 1_2 +
1_2\otimes{\bf \beta}. \sigma + R_{ij} \sigma_{i}\otimes \sigma_{j}].
\end{equation} 
A straightforward calculation then yields 
\begin{eqnarray} 
&t&(\rho; a , {\bf n} ; b,{\bf m}) =\nonumber\\
& &\quad \mu^2\nu^2\big[ (1 + a^2)(1+b^2) +
2 a(1+b^2) {\bf n}.{\bf \alpha} + \nonumber\\ 
& &\qquad 2b(1+a^2) {\bf m} . {\bf \beta} 
+ 4 ab R_{ij} 
n_i m_j\big].
\end{eqnarray}

For a
Werner state ${\bf \alpha} = {\bf \beta} =0$, $R_{ij} = {1 - 4 F \over
3} \delta_{ij}$, and
\begin{eqnarray}
&t&(W(F); a , {\bf n} ; b,{\bf m}) \nonumber\\ 
&{}&\quad =\mu^2\nu^2\left[  (1 + a^2)(1+b^2)
+ {4 \over 3} (1-4F) ab {\bf n}. {\bf m}\right].\label{normalisation}
\end{eqnarray}
Simple algebra then shows that
 $C(\rho_{final})\leq C(\rho)$, which proves that the entanglement of 
formation of a Werner state can never be
increased by local operations on a single copy.  (In fact 
it is possible to show that for any Bell-diagonal state (i.e.
one with $\alpha=\beta=0$), the entanglement 
of formation cannot be increased by local actions on an 
individual copy.)

We note that the above result also shows that local operations and classical 
communication cannot increase the {\em fidelity} of an entangled Werner matrix.
This is because, for Werner matrices, the entanglement of formation is an 
increasing function of the fidelity (although $\rho_{final}$ 
is not necessarily 
of Werner form, it can be randomized, and thus brought into Werner form,
without increasing its entanglement of formation).

The results that we have described above have been obtained by brute force.
However we would like to understand in a deeper way why density matrices
behave differently from pure states as far as their purification is concerned.
By actions on a single copy of any entangled pure state one can extract a 
singlet, with finite probability.  Why can the same thing not be achieved for 
density matrices?  We do not know the complete answer to this question yet.
However can gain some intuition by analyzing the following scenarios.

Consider first the case of a pure state $\psi $ of two spin 1/2 particles.  We 
wish
to obtain a singlet from it.  This can be achieved \cite{BBPS} but only with a 
given probability $P$ of success.  This probability depends on the initial 
state 
$\psi$.   Indeed, the overall average amount of entanglement in the system 
cannot increase, so
the initial entanglement $E_\psi$ and the probability $P$ must satisfy the
inequality
\begin{equation}
P\ E_{singlet}  \leq E_{\psi}.
\end{equation}
Thus if we start from different initial states $\psi$ which are closer and 
closer 
to a non-entangled state, one finds that although one can always obtain a 
singlet, the probability of success must, and indeed does, go  to zero.

Now suppose  
it were the case that purifying density matrices can be achieved in 
a 
similar way, namely
that  a given goal (a given final state) can always be obtained from any 
initial state $\rho$ with some non-zero probability of success.  Specifically, 
let our goal be to obtain a fixed Werner state with fidelity $F_{final}> 
1/2$ (it may be too ambitious to try to obtain a singlet, so we do not assume 
that $F_{final}=1$)
and let us assume that for any initial fidelity $1/2< F< F_{final}$
this can be done.  
Once again if we consider what happens as  our initial state 
tends towards the unentangled Werner state (with $F=1/2$), the probability of 
success {\em must} tend towards zero.
However, as we show below, it turns out that no matter what local actions we 
perform, any possible outcome of the measurement occurs  with a finite 
probability which does not tend to zero as $F\rightarrow 1/2$.  Roughly 
speaking, this is because the non-entangled limit of the family of Werner 
state, 
namely the Werner state with $F=1/2$, is still a mixed state, and the noise 
contained in it does not allow any outcome of any measurement to remain \lq\lq 
silent\rq\rq .
Thus there can be no measurement which could achieve the goal described.

Consider $W_{final}$ to be our fixed goal.  Suppose that it were possible 
to choose actions which allowed one to obtain $W_{final} $ starting from 
$W(F)$.  Then $W_{final}$ would given by
\begin{eqnarray}
& & W_{final} =\nonumber\\
& &\quad {U_A f_A^{a,{\bf n}}\otimes U_B f_B^{b,{\bf m}} \ W(F)\  
 f_A^{a,{\bf n}}  U_A^\dagger \otimes f_B^{b,{\bf m}}  U_B^\dagger
\over t(W(F); a , {\bf n} ; b,{\bf m}) },
\end{eqnarray}
as in (\ref{arbitrary}).
The probability of obtaining $W_{final} $ would then be
equal to the normalization 
$t(W(F); a , {\bf n} ; b,{\bf m})$ which is given in (\ref{normalisation}).
It is straightforward to show that this probability does not go to zero as  
$F\to 1/2$
(excluding the trivial case $\mu $ or $\nu$ equal to zero, in which one  
filters out all the particles). 
Thus  one cannot purify to a fixed output state.

The above argument just shows that mixed states cannot have the same simple 
behavior
as pure states.  More subtle behavior is not ruled out by the argument. For
example it might have  been the case
that an individual Werner state can be purified only in a small range of
fidelities, say $F_{min} < F < F_{final}$,  with $F_{min} >
1/2$.  If this were the case, then as $F \to 1/2$ the probability of
obtaining $W_{final}$ need not tend to zero since as $F\to 1/2$
the state cannot be purified.  The proof given in the first part of
this letter, however, shows that this is not the case.

Finally we note that our expression for the 
eigenvalues of $\rho \tilde \rho$ shows that their ratios
$\lambda_i^2/\lambda_j^2$ 
are
invariant under arbitrary local operations and classical communication 
(excluding randomizations). This therefore provides a
characterization of the equivalence classes of  density matrices under such 
operations.
This  may have important applications because it
provides a simple criterion for distinguishing states whose
entanglement is fundamentally different.
Whether this characterization is complete, i.e.
whether their are additional independent functions of
$\rho$ which are invariant under local  operations and classical communication, 
is still an open
question. Also how to characterize the equivalence of  density matrices under
local operations, classical communication {\em and} randomization is unknown.
(The invariants of  multi-particle 
entangled states under local {\em
unitary} operations have been discussed in \cite{LP,LPS}).

%\bigskip

\noindent{\bf Acknowledgments}
We are very grateful to  Colin Sparrow for his help at an early stage in this
work.  S.P. warmly acknowledges very useful discussions with P.K.~Aravind.

%\vspace{-.5cm}

\end{multicols}

\end{document}